\documentclass[prd,article,nofootinbib,twocolumn,preprintnumbers,superscriptaddress]{revtex4}
\usepackage{graphicx,amsfonts,color,comment,amsmath,hyperref,float}
\usepackage{amssymb}	

\usepackage{mathrsfs,amssymb}  
\usepackage{cancel}
\usepackage[normalem]{ulem}

\newcommand{\be}{\begin{equation}}
\newcommand{\ee}{\end{equation}}
\newcommand{\bea}{\begin{eqnarray}}
\newcommand{\eea}{\end{eqnarray}}

\begin{document}

\title{Detecting Asymmetric Dark Matter in the Sun with Neutrinos}

\author{Kohta Murase}
\author{Ian M. Shoemaker}
\affiliation{Department of Physics; Department of Astronomy \& Astrophysics; Center for Particle and Gravitational Astrophysics, The Pennsylvania State University, University Park, PA 16802, USA}

\date{\today}
\begin{abstract}
Dark Matter (DM) may have a relic density that is in part determined by a particle/antiparticle asymmetry, much like baryons. If this is the case, it can accumulate in stars like the Sun to sizable number densities and annihilate to Standard Model (SM) particles including neutrinos. 
We show that the combination of neutrino telescope and direct detection data can be used in conjunction to determine or constrain the DM asymmetry from data. Depending on the DM mass, the current neutrino data from Super-K and IceCube give powerful constraints on asymmetric DM unless its fractional asymmetry is $\lesssim 10^{-2}$. Future neutrino telescopes and detectors like Hyper-K and KM3NeT can search for the resulting signal of high-energy neutrinos from the center of the Sun.  The observation of such a flux yields information on both the DM-nucleus cross section but also on the relative abundances of DM and anti-DM. 

\end{abstract}
\preprint{}


\maketitle


%
\section{Introduction}
Despite their overwhelming ubiquity we do not know the origin of dark matter or of ordinary baryons. In fact, the similarity of their observed cosmological abundances, $\Omega_{DM} \simeq 5 \Omega_{B}$, may suggest a common origin of dark and baryonic matter. In contrast with models of weakly-interacting massive particles (WIMPs), asymmetric DM (ADM) posits that DM possesses a primordial asymmetry in the relative number of particles and anti-particles, $\left(n_{X} - n_{\bar{X}}\right) \neq 0$ (see e.g.~\cite{Petraki:2013wwa,Zurek:2013wia} for reviews).  However, unlike baryons, the relic abundance of $X$ and $\bar{X}$ may not be enormously different. In this case, ADM can produce annihilation signatures that can be searched for in regions of high DM density in so-called indirect searches~\cite{Graesser:2011wi,Lin:2011gj,Bell:2014xta}. 

One such place where DM may be abundant is the solar interior, where it has been trapped via scattering on nuclei~\cite{Press:1985ug,Krauss:1985ks,DeRujula:1985wm,Gaisser:1986ha,Griest:1986yu,Srednicki:1986vj,Gould:1987ju,Gould:1987ir}. This has been previously used to constrain WIMP dark matter~\citep[{e.g.,}][]{Jungman:1995df} using for example Super-Kamiokande~\cite{Choi:2015ara} and IceCube~\cite{Aartsen:2016exj} data. In fact, because ADM annihilation rates are smaller than WIMPs they can accumulate to very large number densities inside the Sun. Even in lieu of annihilation, this DM can impact the Sun by altering the transport of heat in the solar interior.  In fact, some of the early solutions to the solar neutrino problem suggested that ADM accumulation in the Sun could be responsible~\cite{Krauss:1985ks,Gelmini:1986zz}. More recently this concern has been revived due to revised estimates of solar metalicities which appear to render solar models in strong tension with helioseismology data~\cite{Frandsen:2010yj,Lopes:2012af,Vincent:2014jia,Vincent:2015gqa,Vincent:2016dcp}.  These models can be tested by low-threshold direct detection experiments like CRESST-II~\cite{Angloher:2016jsl} and CDMSlite~\cite{Agnese:2015nto}, and as we will argue here, also by neutrino telescopes like IceCube.

In this paper we consider the impact of annihilating ADM on the Sun and the prospects for its detection at neutrino telescopes. This may even allow for a determination of the dark asymmetry from data, and is one of the view avenues for doing so.  The crucial insight is that in this case the flux of neutrinos from the Sun coming from ADM annihilation, $\Phi_{\nu} \propto r_{\infty} \sigma_{SD}$, where $\sigma_{SD}$ is the cross section on nuclei and $r_{\infty} \equiv n_{\bar{X}}/n_{X}$, taking $\bar{X}$ to be the sub-dominant species. The degeneracy between the fractional asymmetry and the cross section can be broken with a positive detection at a future direct detection experiment where the number of events simply scales as $\sigma_{SD}$. Thus a detectable signal in both direct detection and neutrino telescopes can be used in conjunction to reveal the presence of a dark asymmetry.

\begin{figure}[b]
\begin{center}
 \includegraphics[width=.45\textwidth]{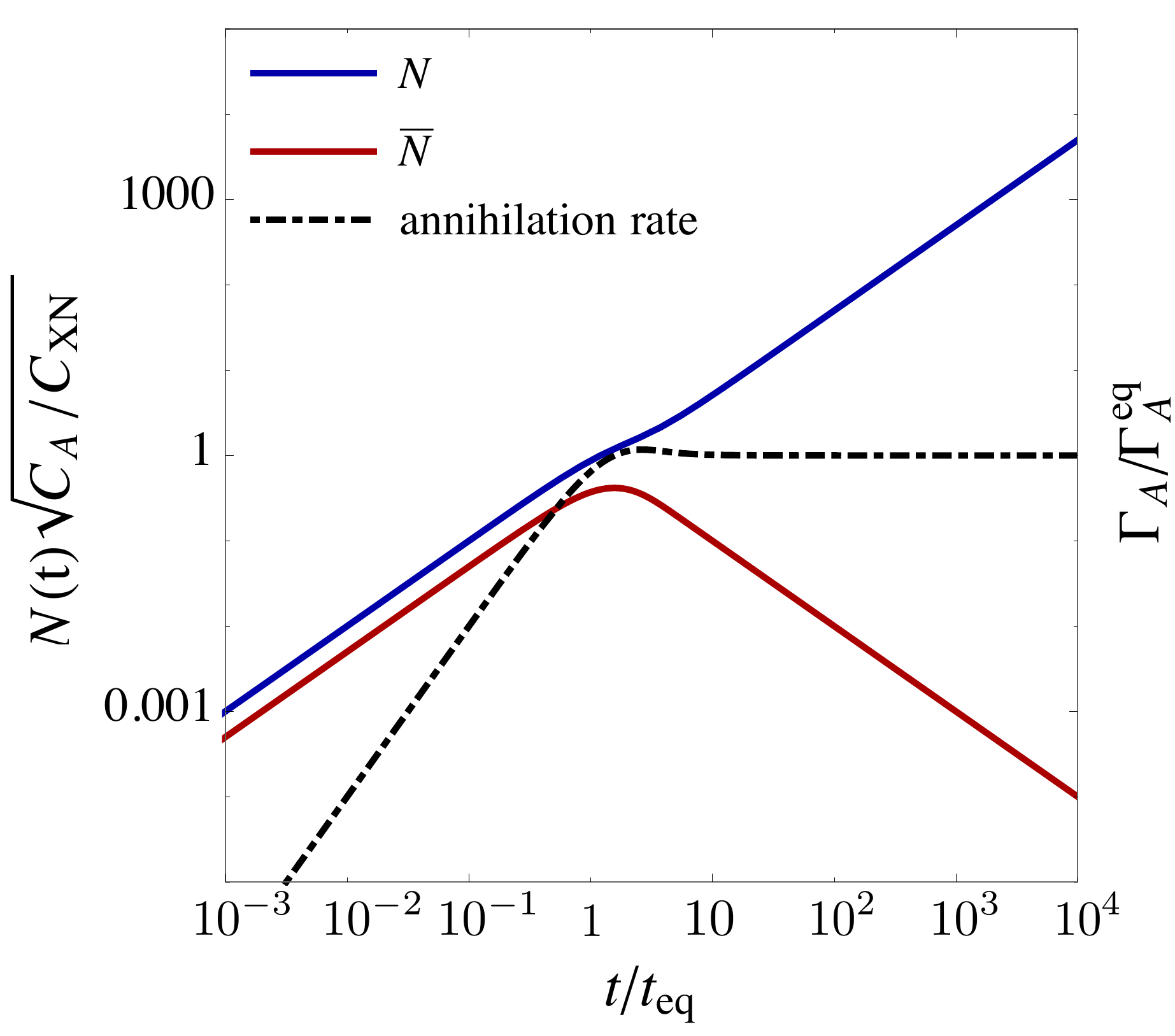} 
\caption{An example of the typical evolution of DM, $X$ and anti-DM, $\bar{X}$ in the Sun as a function of time (in this case {$r_{\infty}=0.5$}). The time at which the annihilation rate reaches equilibrium is $t_{{\rm eq}} \equiv 1/\sqrt{C_{A}C_{C}}$. The final annihilation rate is well-approximated by the analytic estimate $\Gamma_{A}^{{\rm eq}} = r_{\infty} C_{C}$ (see. Eq~(\ref{annrate})).} 
\label{windows}
\end{center}
\end{figure}

\begin{figure}[b]
\begin{center}
 \includegraphics[width=.49\textwidth]{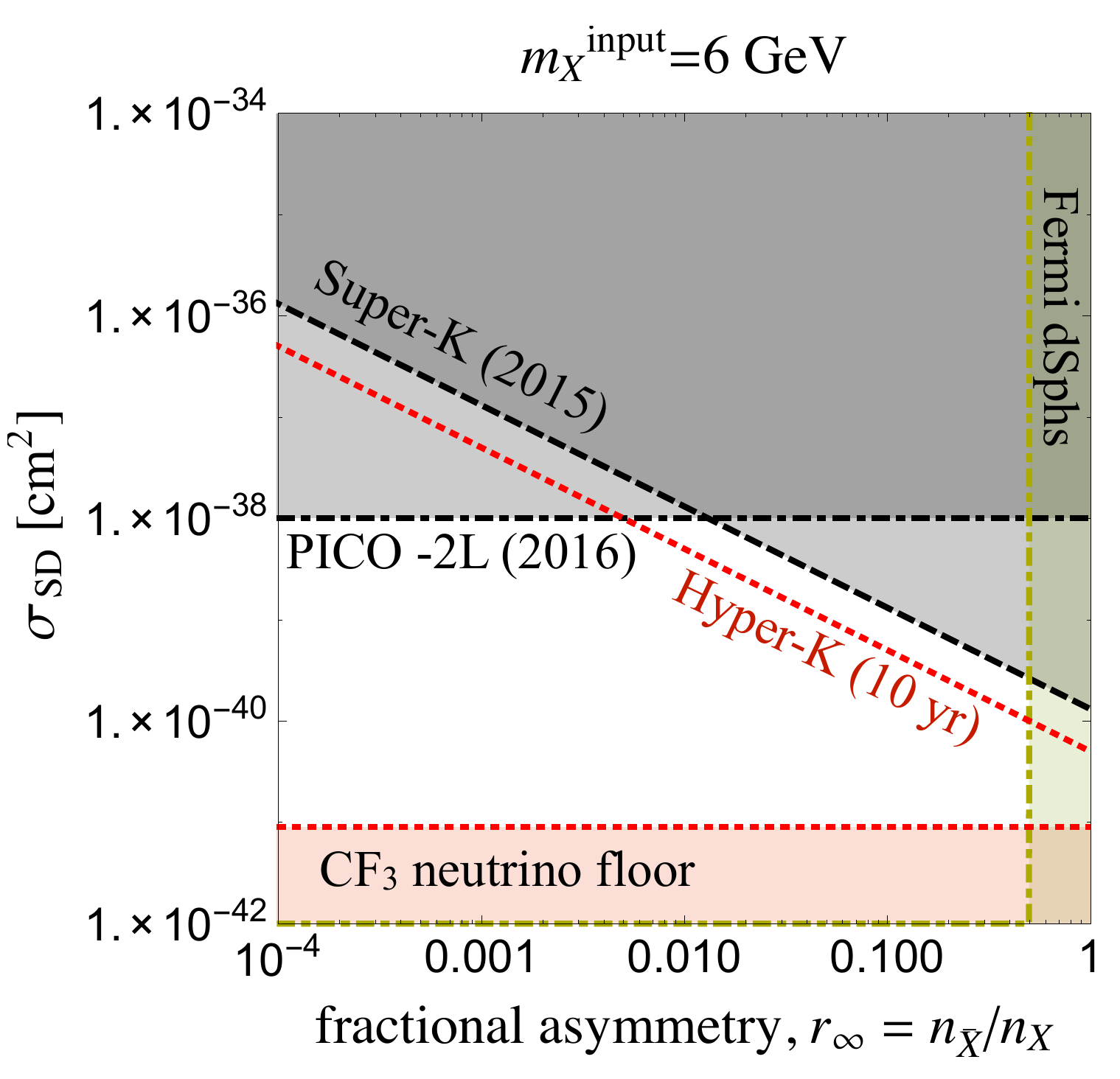} 
\caption{Here we summarize current constraints on light ADM from neutrino telescopes~\cite{Choi:2015ara}, PICO-2L~\cite{Amole:2016pye}, and from Fermi gamma-ray data~\cite{Bell:2014xta}. We also include for reference the neutrino floor background for a $\rm{CF}_{3}$ experiment (e.g. PICO)~\cite{Ruppin:2014bra} and a 10 yr projection of Hyper-K~\cite{HyperKDM}. }
\label{current}
\end{center}
\end{figure}

The remainder of this paper is organized as follows. In Sec.~\ref{sec:cap} we study the way in which a asymmetry modifies the accumulation and annihilation of DM in the solar interior.  We find that although the sub-dominant species (e.g. say $\bar{X}$) eventually starts to be depleted from the Sun the annihilation rate is constant and observable for a wide range of cross sections and asymmetries. We show that these features allow for neutrino telescopes to place competitive constraints on the fractional asymmetry, and may allow for its detection in future data. In Sec.~\ref{sec:model} we consider the implications of these results on the parameter space of an illustrative axial vector simplified model, and in Sec.~\ref{sec:conc} we summarize and conclude.

\section{Solar Capture and Annihilation of DM}
\label{sec:cap}
The cosmological abundance of DM and anti-DM is parameterized by the quantity $r_{\infty} \equiv n_{\bar{X}}/n_{X}$. Thus the evolution of the relative abundances of DM species in astronomical objects is
\bea
\dot{N} &=& C_{C} -C_{A} N \bar{N} \label{eq1}\\
\dot{\bar{N}} &=& r_{\infty} C_{C} -C_{A} N \bar{N}\label{eq2}
\eea
where $C_{A}$ controls the annihilation rate and $C_{C}$ the capture rate from scattering on nuclei, and their specific functional form depends is model-dependent. {For DM masses above a few GeV the effects of evaporation can be ignored~(see e.g. \cite{Busoni:2013kaa}). }

For simplicity, we make rather conventional model assumptions and assume that DM-nuclear scattering is elastic and momentum-independent. This is the most studied case in the literature, though both inelastic~\cite{Nussinov:2009ft,Menon:2009qj,Taoso:2010tg,Blennow:2015hzp} and momentum-dependent~\cite{Vincent:2014jia,Catena:2015uha,Catena:2015iea,Vincent:2015gqa} scattering have both been examined. We shall furthermore assume that the interactions are spin-dependent. This amounts to assuming a specific model like axial-vector mediated interactions, or in the language of effective field theory (EFT) an operator of the form $\mathcal{O} = S_{X} \cdot S_{N}$. We leave a more systematic study of the contributions from the various non-relativistic scattering operators in the context of ADM for future study. 

Under this model assumption, the capture rate can be well-approximated up to form factors by~\cite{Rott:2011fh} 
\be
 C_{C} \simeq 2.3 \times 10^{26}~{\rm s}^{-1}~\left(\frac{\sigma_{SD}}{10^{-38}~{\rm cm}^{2}}\right)~\left(\frac{10~{\rm GeV}}{m_{X}}\right)^{2}.
 \label{eq:cap}
\ee
Notice that with the definition of the capture rate in Eqs.~(\ref{eq1}) and (\ref{eq2}), the expressions for capture in the case of WIMPs and for ADM are identical. The distribution of DM in the vicinity of the Sun may be influenced by the gravity of the other planets in the Solar System~\cite{Gould:1990ad,Damour:1998rh,Peter:2009mi,Peter:2009mk,Sivertsson:2012qj}. At present we do not include these effects.

The expression for the annihilation rate $C_{A}$ inside the Sun also follows the form familiar for WIMPs under the assumption that DM follows a thermal distribution~\cite{Griest:1986yu}
\be C_{A} = \langle \sigma_{\rm ann} v_{\rm rel}\rangle \frac{V_{2}}{V_{1}^{2}}
\label{eq:ann}
\ee
where $V_{j} = 2.45 \times10^{27}~\left(\frac{100~{\rm GeV}}{j~m_{X}}\right)^{3/2}~{\rm cm}^{3}$. {We note that in standard WIMP scenario, one simply solves Eq.~(\ref{eq1}) using Eqs.(\ref{eq:cap}) and (\ref{eq:ann}) for the DM capture and annihilation rates respectively, while taking the thermal relic annihilation cross section~$\langle \sigma_{{\rm ann}} v_{{\rm rel}} \rangle \simeq 6 \times 10^{-26}~{\rm cm}^{3}~{\rm s}^{-1}$.}  

In ADM models however one must account for the presence of a nonzero asymmetry in solving the Boltzmann equations for the relic abundances of $X$ and $\bar{X}$.  When the present-day abundance of the sub-dominant species is small (i.e. $r_{\infty} \ll1$),  the required annihilation cross section is~\cite{Graesser:2011wi,Lin:2011gj}
\be 
 \langle \sigma_{{\rm ann}} v_{{\rm rel}}\rangle_{ADM} = \sqrt{\frac{45}{\pi}}\frac{(n+1)x_{f}^{n+1}s_{0}}{\rho_{c}\Omega_{DM}M_{Pl}\sqrt{g_{*}}}~\log \left(\frac{1}{r_{\infty}}\right),
 \ee
 {where $\rho_{c}$ is the critical density, $\Omega_{DM}$ is the DM density in units of $\rho_{c}$, $M_{Pl}$ is the Planck mass, $s_{0}$ is the present day entropy density, $g_{*}$ is the effective number of relativistic degrees of freedom, and $x_{f} \equiv m_{X} /T_{F} \simeq 20$ where $T_{F}$ is the DM freeze-out temperature which only weakly depends on the DM properties~\cite{Graesser:2011wi}. Lastly the integer $n$ characterizes the temperature dependence of the annihilation, $ \langle \sigma_{{\rm ann}} v_{{\rm rel}}\rangle \propto T^{n}$. }
 
 This allows one to rewrite the annihilation rate directly in terms of the fractional asymmetry
 \be 
 C_{A} \simeq 2.3\times10^{-55}~{\rm s}^{-1}~\left(\frac{m_{X}}{10~{\rm GeV}}\right)^{3/2}~\log \left(\frac{1}{r_{\infty}}\right)
 \ee

 \begin{figure*}[t]
\begin{center}
 \includegraphics[width=.45\textwidth]{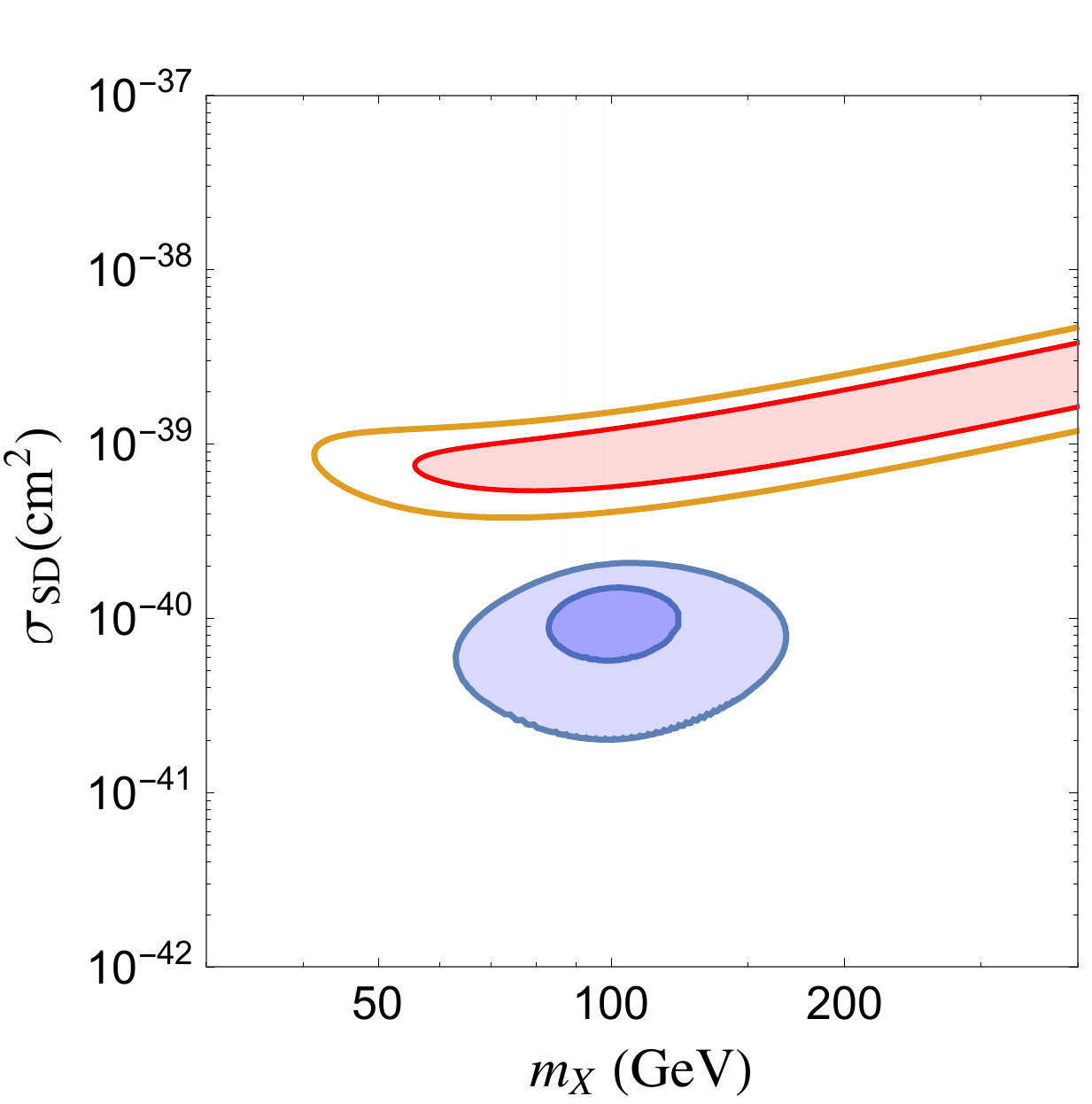} 
 \includegraphics[width=.45\textwidth]{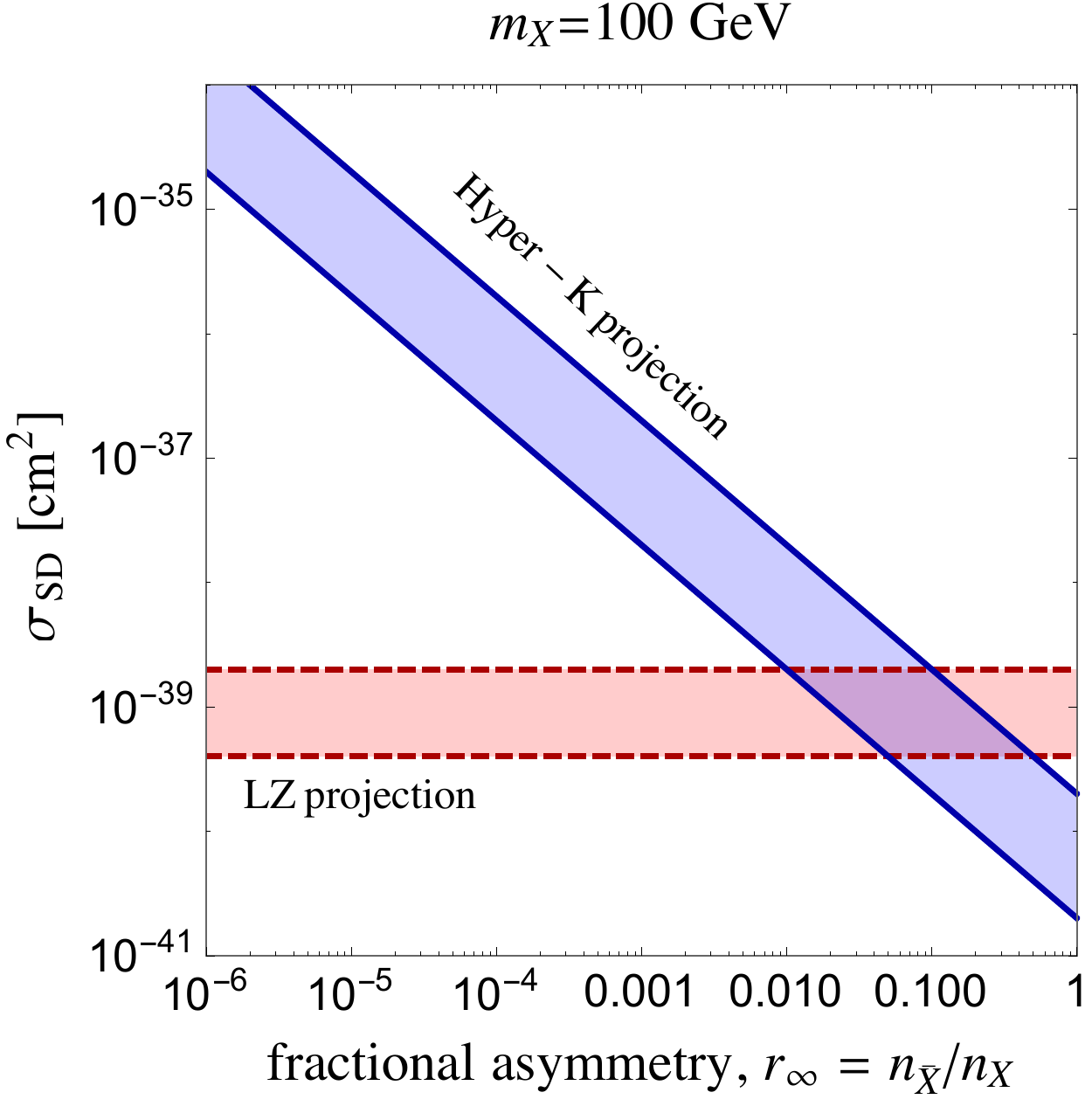} 
\caption{{\it Left}: 1 and 2$\sigma$ best-fit regions for LZ (red) and Hyper-K (blue) mock data assuming pure $\tau \tau$ annihilation. Note that the Hyper-K region {\it incorrectly assumes} a WIMP-like model for the fit (i.e. $r_{\infty} =1$). The strong tension between the LZ and Hyper-K reconstruction motivates the ADM model. Here the input DM parameters are $m_{X} =100$ GeV, $\sigma_{SD} = 10^{-39}~{\rm cm}^{2}$, and $r_{\infty} =0.1$. We have taken 5 Mton-yr exposure for Hyper-K and a $5.6\times10^{5}$ kg-day (5.6 tonne fiducial mass with 1000 days) exposure for LZ in the 6-50 keV energy window~\cite{Akerib:2015cja}. In the {\it right panel} we show the best-fit regions having marginalized over the DM mass. The joint posterior region prefers $r_{\infty} <1$ at more than 3$\sigma$. Note that with 100 GeV DM PICO-2L requires $\sigma_{SD}^{p} < 10^{-39}~{\rm cm}^{2}$~\cite{Amole:2016pye}.}
\label{windows}
\end{center}
\end{figure*}

At the earliest times of stellar history the evolution is dominated by accretion. Thus at these early times ($t < t_{{\rm eq}}$) we have simply
\bea 
N(t)& \simeq& C_{C} t \\
\bar{N}(t) &\simeq & r_{\infty} C_{C}t.
\eea
The importance of the annihilation terms in Eqs.~(\ref{eq1}) and (\ref{eq2}) grows with time, such that at a time 
\begin{eqnarray}
 t_{{\rm eq}} &\equiv& \frac{1}{\sqrt{C_{A}C_{C}}}\nonumber\\
& \simeq&  2.5~{\rm Myr}~\left(\frac{m_{X}}{{\rm GeV}}\right)^{1/4}\left(\frac{10^{-38}~{\rm cm}^{2}}{\sigma_{{\rm SD}}}\right)^{1/2}\frac{1}{\sqrt{\ln\left(1/r_{\infty}\right)}}\nonumber
 \end{eqnarray}
the annihilation rate is in equilibrium with the {\it more abundant} species, $X$. This temporary balancing of capture and annihilation occurs a factor $1/\sqrt{r_{\infty}}$ later for the sub-dominant $\bar{X}$ species. So far the evolution of the solar DM abundance in the ADM and WIMP cases are qualitatively similar. 

Turning now to times beyond $t_{{\rm eq}}$ we will see that the evolution in the ADM case is markedly different than what occurs for WIMPs.  For ease of illustration, in this regime we will assume that the fractional asymmetry is small, $r_{\infty} \ll 1$, such that the linear growth regimes applies for all times to $N$,
\be
N(t) \simeq \sqrt{\frac{C_{C}}{C_{A}}} + \left(t-t_{{\rm eq}} \right) C_{C} \rightarrow N(t)\simeq C_{C} t.
\ee
The final expression is obtained in the limit of $t\gg t_{\rm eq}$.
Then the evolution of the sub-dominant species is easy to track
\be 
\dot{\bar{N}} \simeq r_{\infty} C_{C} -(t/t_{{\rm eq}}^{2})\bar{N}~\label{eqNbar}
\ee
Now Eq.~(\ref{eqNbar}) can be solved to yield 
\be
\bar{N}(t) = e^{-t^{2}/2t_{{\rm eq}}^{2}}\left( \beta + \sqrt{\frac{\pi}{2}} N_{{\rm eq}} r_{\infty} ~{\rm Erfi}\left( \frac{t}{\sqrt{2} t_{{\rm eq}}}\right) \right) \label{asymp}
\ee
where $\beta$ is determined from the requirement that $\bar{N}(t_{{\rm eq}}) = r_{\infty} \sqrt{\frac{C_{C}}{C_{A}}}$ but is irrelevant in what follows. Using the asymptotic behavior of the ${\rm Erfi}(x)$ function, ${\rm Erfi}(x) \rightarrow \frac{1}{\sqrt{\pi}}\frac{e^{x^{2}}}{x}$, it can be shown that the solution in Eq.~(\ref{asymp}) asymptotically approaches~({cf.~\cite{Griest:1986yu} for neutrino dark matter})

\be 
\bar{N}(t) \simeq \frac{r_{\infty}}{C_{A}}~ t^{-1}, ~~~~~{\it as}~t\rightarrow \infty.
\ee
Thus despite the fact the evolution of neither particle species equilibrates, the annihilation rate does reach a steady value:
\bea \Gamma_{{\rm ann}} &\equiv& C_{A} N \bar{N} \\
& \simeq& r_{\infty} C_{C} \label{annrate}
\eea
Therefore the solar abundance of ADM renders itself testable by constraints on high-energy neutrino fluxes from the Sun. 

An additional important difference that ADM has compared to WIMPs is the possibility that we are observing the Sun today so far after $t_{{\rm eq}}$ that the $\bar{X}$ abundance vanishes, and there ceases being any significant annihilation. Defining this time as $N(t_{\bar{X}}) = 1$ we find 
\be
t_{\bar{X}} = 4.4\times 10^{48}~{\rm yr}~\left(\frac{{\rm GeV}}{m_{X}}\right)^{3/2} \frac{r_{\infty}}{\log (1/r_{\infty})}.
\ee
We therefore conclude that for GeV-scale ADM $t_{\odot} < t_{\bar{X}}$ is achieved as long as $r_{\infty} \gtrsim 10^{-39}$, which is comfortably within the range of fractional asymmetries of interest.

Finally we compute the number of signal events at the detector. We assume an idealized detector such that the rate depends only on physical quantities (i.e. the analysis acceptance is implicitly assumed to be perfect). 
{Our calculation of the event rates are given in Appendices.}

\subsection{Discussion of Results}
{Now we turn to the implications of these results. In Fig.~\ref{current} we display current limits on a low-DM example in the cross section - fractional asymmetry plane. In view of Eq.~(\ref{annrate}), solar neutrino flux limits can be recast as limits on the quantity $r_{\infty} \sigma_{SD}$. The cross section is also itself directly constrained by direct detection data while the fractional asymmetry can be constrained by indirect detection data~\cite{Graesser:2011wi,Lin:2011gj,Bell:2014xta}. We see that present Super-K data~\cite{Choi:2015ara} already cuts into previously unconstrained parameter space. Moreover, Hyper-K will be able to further constrain the fractional asymmetry down to the $r_{\infty}\lesssim 5\times 10^{-3}$ level for cross sections near the present PICO-2L limits~\cite{Amole:2016pye}. Note that sensitivity to low-mass DM will be improved with PINGU for example~\cite{Aartsen:2014oha}. }

{Next we examine the kind of improvement that can be reached in the near future if a signal of solar DM annihilation is found. The technical details of our fit are provided in the Appendix. First we consider what the combined data from future direct detection (e.g. LZ) and neutrino telescope (e.g. Hyper-K) may reveal. As an example, we simulate mock data coming from 100 GeV ADM particle annihilating purely into $\tau \tau$ final states with a fractional asymmetry of $r_{\infty} = 0.1$. For illustration, in left panel of Fig.~\ref{windows} we assume such data would initially be fit under the {\it incorrect} assumption of WIMP DM. The resulting disagreement between the LZ and Hyper-K contours is striking and shows that when faced with such data the WIMP interpretation would be clearly lacking.} 

{Nevertheless the offset in the inferred best-fit cross sections can be used to estimate the fractional asymmetry. In the right panel of Fig.~\ref{windows} we show what the best-fit regions after marginalizing over the DM mass. We find that the symmetric WIMP interpretation (i.e. $r_{\infty}=1$) can be rejected at $> 3\sigma$, and can therefore be used to determine the fractional asymmetry from data. }

\begin{figure}[t]
\begin{center}
 \includegraphics[width=.49\textwidth]{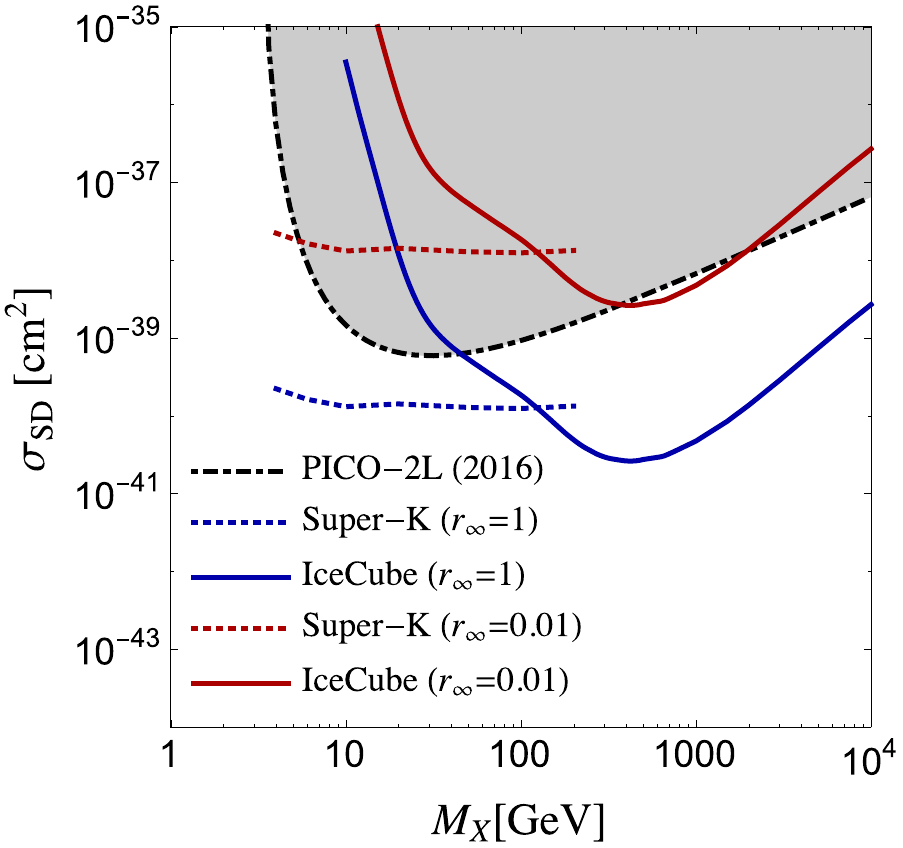} 
\caption{For illustration the current IceCube~\cite{Aartsen:2016exj} and Super-K~\cite{Choi:2015ara} limits on $\tau\tau$ annihilation are rescaled according to Eq.~(\ref{eq:ann}) by $r_{\infty}$ until neutrino telescopes and direct detection limits (PICO-2L~\cite{Amole:2016pye}) are comparable. We see that neutrino telescopes can probe down to $r_{\infty} \simeq 0.01$ with present data.}
\label{compare}
\end{center}
\end{figure}

Finally, it is important to highlight that for sufficiently small fractional asymmetry $r_{\infty}$ the sensitivity of neutrino telescopes will be weaker than direct detection. This is illustrated pictorially in Fig.~\ref{compare}, where we see that neutrino telescopes provide meaningful constraints on ADM down to $r_{\infty} \simeq 0.01$ for $\tau \tau$ annihilation. The analogous exercise for $bb$ annihilation reveals that present data only allows us to probe down to $r_{\infty} \simeq 0.7$. Note that although we have taken PICO-2L for illustration, PICO-60~\cite{Amole:2015pla} has slightly stronger limits on the DM-proton spin-dependent cross section for $\gtrsim 50$ GeV masses (and both experiments are stronger than the current limits from LUX~\cite{Akerib:2016lao}).

Throughout our analysis we have assumed neutrino-rich annihilation channels. If instead these branching ratios are small or vanish, then the limits from neutrino telescopes will be correspondingly weakened. Lastly note that we have not investigated the detailed sensitivity of KM3NeT~\cite{Adrian-Martinez:2015rtr}, but we anticipate that it will provide additional constraints on solar ADM. 

\begin{figure}[t]
\begin{center}
 \includegraphics[width=.49\textwidth]{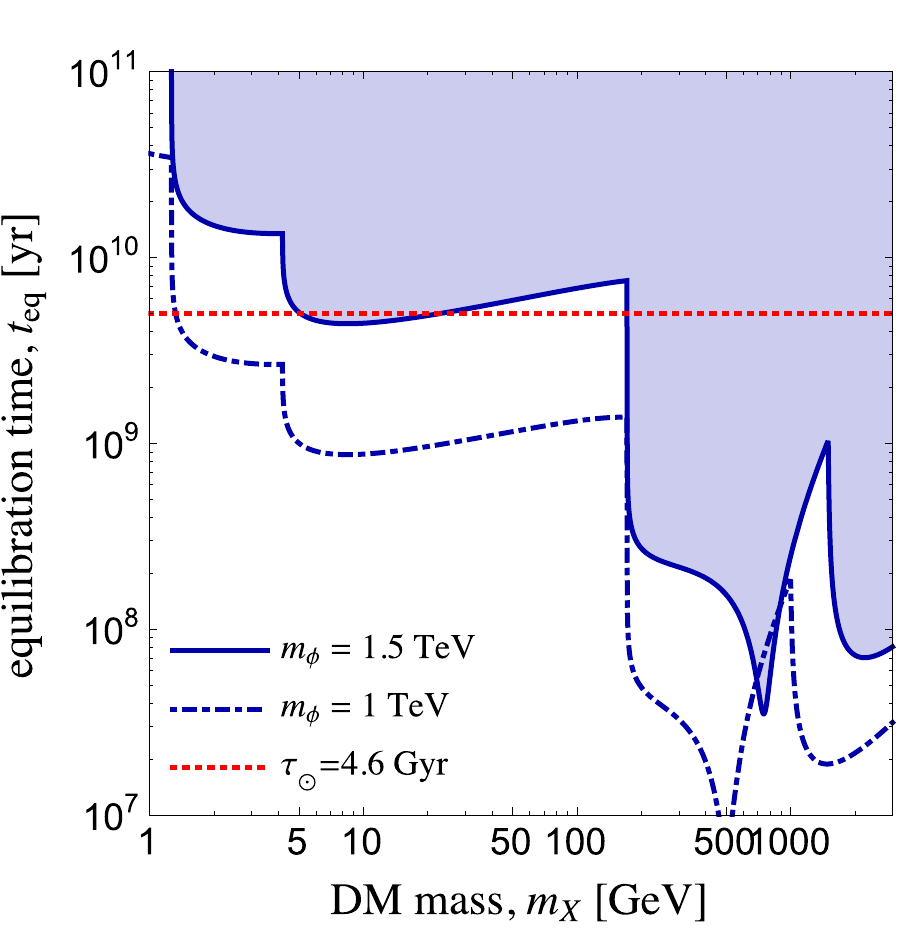} 
\caption{We see that equilibrium is reached for a range of fractional asymmetries for TeV scale mediators. The time at which the annihilation rate reaches equilibrium is $t_{{\rm eq}} \equiv 1/\sqrt{C_{A}C_{C}}$. }
\label{fig:equil}
\end{center}
\end{figure}

\section{Model Implications}
\label{sec:model}
The main results of this paper apply to any ADM model with sizable cross sections and fractional asymmetries. Let us illustrate the impact of this constraint by briefly examining the implications for the following simplified model with an axial vector mediator 
\be 
\mathscr{L} \supset \left(g_{f} \bar{f} \gamma^{5}\gamma^{\mu} f + g_{X}\bar{X} \gamma^{5}\gamma^{\mu} X\right) \phi_{\mu},
\ee
where $\phi$ is a vector mediator field. The $s$-channel annihilation cross section is to leading order
\be
\langle \sigma v \rangle_{\bar{X}X \rightarrow \bar{f}f} \simeq \frac{N_{f}m_{X}^{2}}{2\pi} \frac{ \frac{m_{f}^{2}}{m_{X}^{2}} + v^{2} }{\left(m_{\phi}^{2} - 4m_{X}^{2} \right)^{2}+\Gamma_{\phi}^{2} m_{\phi}^{2}}~\sqrt{1-\frac{m_{f}^{2}}{m_{X}^{2}}}
\label{eq:model}
\ee
where $N_{f}$ is a color factor that is 3 for quarks and 1 for leptons, and $\Gamma_{\phi}$ is the total decay width of the mediator. As we will see, the factor of $(m_{f}^{2}/m_{X}^{2} + v^{2})$ plays a crucial role in this model by making the annihilation rate strongly dependent on the DM mass. Note that in our numerical implementation we use the complete expression~\cite{Berlin:2014tja,Heisig:2015ira}, and include the annihilation channel to two mediators (which dominates for $m_{X} > m_{\phi}$).

By assuming couplings to SM fermions, this simplified model is constrained by the complementary searches at colliders (via $\bar{f} f \rightarrow \bar{X}X$), direct detection (via $ X f \rightarrow X $) and indirect detection searches (via $\bar{X} X \rightarrow \bar{f}f$).  Of course compared to WIMPs only the last of these search categories have modified event rates for ADM.

The limits from monojets at high-energy colliders like the Tevatron and LHC are quite strong for kinematically accessible DM~\cite{Aaltonen:2012jb,ATLAS:2012zim,Khachatryan:2014rra}. Indeed they are stronger than those offered by current direct detection sensitivity (e.g. LUX) for DM masses $\lesssim 100$ GeV with a weak dependence on the couplings $g_{X}, g_{f}$. 

However for higher DM masses, the rates at neutrino telescopes can become sizable~\cite{Heisig:2015ira}. In this axial vector model, this is largely a result of the cross section scaling, $\langle \sigma v \rangle \propto m_{f}^{2}/m_{X}^{2}$. For DM masses a bit above the top quark threshold, the annihilation channel $\bar{X}X \rightarrow \bar{t}t$ becomes kinematically open and dominates the total annihilation rate. The detectability is heightened further above top threshold since the spectrum of 
neutrinos from $\bar{b}b$ annihilation is much less constrained. 

{To achieve large rates at neutrino telescopes we must be currently at times larger than $t_{eq}$. We display in Fig.~\ref{fig:equil} the equilibrium time $t_{eq}$ as a function of the DM mass for two choices of the mediator mass. As can be seen the value for $t_{eq}$ depends sensitively on the DM mass. Again, this arises because of the helicity suppressed annihilation in Eq.~(\ref{eq:model}) and the fact that $\bar{X} X \rightarrow \phi \phi$ dominates once $m_{X} \gtrsim m_{\phi}$. }

Using Eq.~(\ref{annrate}) and comparing the limits from LUX's 2016 spin-dependent limits~\cite{Akerib:2016lao} and IceCube's recent 3 year data~\cite{Montaruli:2015six} we find that our results on solar ADM imply that IceCube's current limit is stronger than LUX for $r_{\infty} \gtrsim 0.1$ and $m_{X} \gtrsim 200$ GeV. Note moreover that most of the parameter space in this model cannot be constrained for low DM masses because the the equilibration time is much longer than the age of the Sun, see Fig.~\ref{fig:equil}.

Note that although Super-K has much stronger limits for some annihilation modes at low masses as compared to IceCube. {The constraints can be further improved by Hyper-K~\cite{HyperKDM}.} However, as noted above these limits will not be strong for low DM masses in this specific model because of the strong suppression of the total annihilation cross section for low $m_{X}$.

\section{Conclusions}
\label{sec:conc}

We have studied the sensitivity of future neutrino telescopes to the presence of asymmetric dark matter in the Sun. Importantly this can yield information on the asymmetry, or in the case of a non-detection significantly constrain it. This enables us to probe the presence of ADM in the solar interior, which is a useful complement to the sensitivity offered by helioseismology data~\cite{Frandsen:2010yj,Lopes:2012af,Vincent:2014jia,Vincent:2015gqa}. Neutrino observations can offer additional means of testing the DM hypothesis in the presence of a positive signal and as a tool for learning about the microphysical properties of DM. {We stress that this is only a first study in this direction and that the sensitivity to low-mass DM annihilation offered by stopped {pions}~\cite{Rott:2012qb} and kaons~\cite{Rott:2015nma} may significantly strengthen the conclusions reached here.}

Lastly one should view the solar sensitivity to ADM more broadly in the context of the other existing astrophysical constraints on ADM. 
These can be very qualitatively different depending on the spin of the ADM particle and the fractional asymmetry. 
For example, bosonic ADM has been ruled out for a range of masses and cross sections $\sigma > 10^{-50}~{\rm cm}^{2} ({\rm GeV}/m_{X})$~\cite{Kouvaris:2011fi}, though repulsive DM self-interactions can substantially weaken this limit~\cite{Bell:2013xk}. 
{Intriguingly, the gravitational collapse induced by ADM accumulation may account for the lack of old millisecond pulsars in the galactic center of the Milky Way~\cite{Bramante:2014zca}. Also, the ADM induced collapse of a star can also be probed with future gravitational wave signals~\cite{Kurita:2015vga}.
Importantly, however, the presence of even a small amount of annihilation can also render such considerations nearly entirely unconstraining~\cite{Bramante:2013hn}.  
}

\section*{Acknowledgements}
{We are very grateful to Carsten Rott for helpful comments and a careful reading of the manuscript.} The authors would like to thank the Pennsylvania State University for support. As a IGC Fellow I. M. S. would also like to thank the Institute for Gravitation and the Cosmos their support.

\section*{Appendix I: {Event Rates in Indirect DM Detection by Neutrino Detectors}}
We follow the method outlined in Ref.~\cite{Rott:2011fh} for obtaining signal and background rates. First we must include the effect of neutrino oscillations as well as the model-dependent branching ratios and corresponding neutrino spectra for each given annihilation channel. We will mostly examine the up-going muon channel for solar DM limits. 

Since the detection properties of muon neutrinos and muon anti-neutrinos are not very different in this search, we will simply sum their contributions and define the terrestrial frame neutrino spectrum as 
\be
n_{\nu_{\mu}} = \sum_{i} P(i\rightarrow \mu) \sum_{f} {\rm Br}(\bar{X}X \rightarrow \bar{f}f)~\frac{dN_{i}^{f}}{dE}
\label{eqflavor}
\ee
where ${\rm Br}(\bar{X}X \rightarrow \bar{f}f)$ controls the model-dependent DM branching ratios, the index $j$ runs over neutrino flavor, $P_{j\rightarrow \mu}(E)$ represents the probability that a neutrino produced as flavor $j$ oscillates into a $\mu$ neutrino, and $\frac{dN_{i}^{f}}{dE}$ is the contribution of $i$ flavor neutrinos to annihilation channel $f$.

Now the differential event rate at the detector via
\be
\frac{dN^{{\rm sig}}_{\nu_{j}}}{dE}= Tn ~\frac{\Gamma_{A}}{4 \pi D_{\odot}^{2}}\sigma^{{\rm CC}}_{\nu_{j}N}(E) n_{\nu_{j}}^{f}
\ee
where $ D_{\odot}$ is the average Earth-Sun distance, $T\cdot n$ is the exposure, $\sigma^{{\rm CC}}_{\nu_{j}N}$ is the charged current neutrino-nucleus cross section, the annihilation rate $\Gamma_{A}$ is well-aproximated by Eq.~(\ref{annrate}),  and the spectrum of neutrinos $ n_{\nu_{\mu}}$ is given by Eq.~(\ref{eqflavor}). We use the spectra for various annihilation channels from Ref.~\cite{Cirelli:2005gh,Baratella:2013fya}, {which gives the spectra at the detector including neutrino oscillations, and matter effects in the Sun and Earth}.

We also mock up the background following~\cite{Rott:2011fh}
\be 
\frac{dN^{{\rm bkg}}}{dE} =Tn~\sigma_{\nu_{\mu}}^{{\rm CC}}(E) \phi_{\nu_{\mu}}^{{\rm atm}}(E) \int_{0}^{\Psi(E)}2\pi \cos(\theta) d\theta.
\ee
where $\Psi(E)$ is the energy-dependent opening angle centered on the Sun.

\section*{Appendix II: {Event Rates at Direct Detection}}

We compute the event rate at direct detection experiments in the following manner~(see e.g.~\cite{Jungman:1995df,Smith:1988kw} for reviews), 
\bea \frac{dR}{dE_{R}} &=& \frac{\rho_{\odot}}{m_{N}m_{X}}\left\langle v\frac{d \sigma}{dE_{R}} \right\rangle \\
&=& \frac{\rho_{\odot}}{m_{N}m_{X}} \int_{v_{min}(E_{R})}^{\infty} d^{3} v~ v f(\vec{v}+\vec{v}_{e}(t)) \frac{d\sigma}{dE_{R}}\nonumber, 
\eea
where $\mu_{N}$ is the DM-nucleus reduced mass, $\vec{v}_{e}(t)$ is the velocity of the laboratory with respect to the rest frame of the galaxy, $f(v)$ is the local velocity distribution of DM, and $\rho_{\odot}$ the local DM density.  $v_{min} (E_{R})$ is the minimum DM velocity to produce a nuclear recoil of energy $E_{R}$ which for elastic scattering, is $v_{min} (E_{R}) = \sqrt{m_{N} E_{R}/2\mu_{N}^{2}}$. We follow the convention by reporting results assuming a Maxwell-Boltzmann distribution. 
We consider spin-dependent scattering on the proton, for which the scattering cross section is 
\be 
\sigma_{SD} = \frac{4}{3} \left(\frac{J+1}{J}\right) \frac{\mu_{N}^{2}}{\mu_{p}^{2}} \langle S_{p}\rangle^{2} \sigma_{p}^{SD}
\ee
where $\mu_{p}$ is the proton-DM reduced mass, $J$ is the total nuclear spin, and $\langle S_{p}\rangle$ is the nuclear spin expectation value of the proton group. For Xenon these values are $\langle S_{p} \rangle = 0.010$ for ${}^{129}$Xe ($J=1/2$) and $\langle S_{p} \rangle =- 0.009$ for ${}^{131}$Xe ($J=3/2$)~\cite{Giuliani:2005bd}.

For the LZ projection in the main body of the text we took a background-free $5.6\times 10^{5}~{\rm kg}$-${\rm day}$ exposure over an energy range 6-50 $\rm{keV}$ consistent with Ref.~\cite{Akerib:2015cja}. Note that the results presented in Fig.~\ref{windows} assume the Maxwell-Boltzmann distribution, though astrophysical uncertainties can lead to additional degeneracies in determining the DM mass and cross section from data~(e.g.~\cite{Friedland:2012fa}).

\bibliographystyle{JHEP}

\bibliography{nu}

\end{document}